\documentstyle[psfig]{aastex}

\slugcomment{to appear in the Astronomical Journal} 
 
\begin{document}
 
\title{High resolution imaging of molecular line emission from 
high redshift QSOs} 
 
\author{C. L. Carilli}
\affil{National Radio Astronomy Observatory, P.O. Box O, Socorro, NM,
87801, USA \\
ccarilli@nrao.edu}
\author{Kataro Kohno}
\affil{Nobeyama Radoi Observatory, Nagano, Japan}
\author{Ryohei Kawabe}
\affil{National Astronomical Observatory of Japan}
\author{Kouji Ohta}
\affil{Department of Astronomy, Kyoto University, Japan}
\author{C. Henkel \& Karl M. Menten}
\affil{Max-Planck-Institut f{\"u}r Radioastronomie, Auf dem H{\"u}gel 69,
Bonn, D-53121, Germany \\}
\author{M.S. Yun}
\affil{Astronomy Department, University of Massachusetts, Amherst, MA} 
\author{A. Petric}
\affil{New Mexico Institute of Mining and Technology}
\author{Yoshinori Tutui}
\affil{Institute of Astronomy, University of Tokyo, Japan}
 
\begin{abstract}

We present moderate (1$''$) and high resolution (0.2$''$) observations
of the CO (2--1) emission at 43 GHz, and radio continuum emission at
1.47 GHz, from the $z = 4.7$ QSO BRI 1202--0725 and the $z = 4.4$ QSO
BRI~1335--0417 using the Very Large Array.  The moderate resolution
observations show that in both cases the CO emission is spatially
resolved into two components separated by 1$''$ for 1335--0417 and
4$''$ for 1202--0725.  The high resolution observations show that each
component has sub-structure on scales $\sim 0.2''$ to 0.5$''$, with
intrinsic brightness temperatures $\ge 20$ K.  The CO
ladder from (2-1) up to (7-6) suggests a high kinetic temperature for
the gas ($\rm T_{kin} \simeq 70$ K), and a high column density (N(H$_2$)
$\simeq 10^{24}$ cm$^{-2}$). In both sources the continuum-to-line ratio:
${{L_{\rm FIR}}\over{L'_{\rm CO(1-0)}}} \simeq 335$.
All these characteristics (brightness temperature,
excitation temperature, column density, and continuum-to-line ratio)
are comparable to conditions found in low redshift, ultra-luminous
nuclear starburst galaxies.  We find that the CO emitting regions in
1202--0725 and 1335--0417 must be
close to face-on in order to avoid having the gas mass 
exceed the gravitational mass, implying perhaps unreasonably
large rotational velocities. While this problem is mitigated
by lowering the CO luminosity-to-H$_2$ mass conversion
factor (X),  the required X values become
comparable to, or lower than,  the minimum values dictated by
optically thin CO emission.  We considered the possibility
of magnification by gravitational lensing in order to reduce the
molecular gas masses. 

\end{abstract}
 
\keywords{radio continuum: galaxies --- infrared: galaxies ---
galaxies: active, distances and redshifts, starburst, evolution, radio
lines}  

\section {Introduction}

The recent discoveries that the large majority of spheroidal galaxies
in the nearby universe contain massive black holes, and that the black
hole mass correlates with the mass of the spheroid  of the parent
galaxy, have led to the hypothesis of co-eval formation of massive
black holes and galaxy spheroids, perhaps occurring in merging galaxies
at high redshift (Richstone et al. 1998; Gebhardt et al. 2000;
Ferrarese \& Merritt 2000; 
Franceschini et al. 1999, Blain et al. 1999, Kauffmann \& Haenelt
2000).  We have undertaken an extensive study of high redshift QSOs in
order to address the interesting question of co-eval black hole and
spheroidal galaxy formation. The large samples of high $z$ QSOs coming
from wide field surveys such as the Sloan Digital Sky Survey (Fan et
al. 2001), and the Digitized Palomar Sky Survey (DSS; Djorgovski et
al. 1999) have greatly facilitated the study of high redshift QSOs. Our
studies involve: (i) searches for thermal dust emission at
mm-wavelengths from a large sample of $z > 3.7$ QSOs (Carilli et
al. 2001a; Omont et al. 2001), (ii) high resolution imaging at
cm-wavelengths of the non-thermal radio continuum emission from these
sources (Carilli et al. 2001a, Carilli et al. 2001b), and (iii)
observations of the CO line emission from selected sources with large
infrared luminosities, as inferred from the mm-continuum observations
(Carilli, Menten, \& Yun 1999).

The interesting result from the mm-continuum surveys of high redshift
QSOs is that 30$\%$ of the sources are detected in surveys with flux
density limits of 1 to 2 mJy at 250 GHz (Carilli et al. 2001a; Omont
et al. 2001), corresponding to infrared luminosities $\ge 10^{12}$
L$_\odot$ and dust masses $\ge 10^{8}$ M$_\odot$.  Radio continuum
studies of these sources show that the mm-to-cm spectral energy
distributions (SEDs) of most of the sources are consistent with the
radio-to-far IR correlation found for nearby star forming galaxies
(Condon 1992; Carilli 2001a, b; Yun et al. 2000). If the dust is
heated by star formation, the implied star formation rates are of
order 10$^3$ M$_\odot$ year$^{-1}$.  On the other hand, the global
SEDs for these sources from cm-to-optical wavelengths are not out of
the range defined by lower redshift, lower luminosity QSOs (Carilli et
al. 2001a,b; Sanders et al. 1989), and hence the case for dust-heating
by star formation in these sources is by no means secure.

Searches for CO line emission from 
dust emitting QSOs at $z > 4$ have resulted in 
the detection of molecular line emission from four sources to date: 
1202--0725, 1335--0415, 0952--0115, and 2322+1944 (Ohta
et al. 1996; Omont et al. 1996b; Guilloteau et al. 1997, 1999; Cox et
al. 2001). The implied gas masses are large ($\ge 10^{10}$ M$_\odot$),
leading some to speculate that star formation may 
be inevitable  (Omont et al. 2001).
Moreover, in at least one case (1202--0725) the molecular line
and mm continuum emission is known to be spatially extended, 
with a bright emission region well separated from the optical QSO. 

For sources at $z \ge 3.9$ the CO(2-1) transition is redshifted into
the  43 GHz band of the Very Large Array (VLA), thereby allowing 
for sensitive observations to be made of
the low order transitions from these sources.
The VLA also allows for sub-arcsecond angular resolution observations
to be made, which are critical to the physical interpretation of
the systems. 

In this paper we present high resolution observations of the CO(2-1)
emission from 1202--0725 at $z = 4.7$ and from 1335--0415 at $z =
4.4$, along with radio continuum observations at 1.4 GHz of
1202--0725. These data are combined with observations of higher order
CO transitions in an attempt to understand the physical conditions in
the molecular line emitting regions of high $z$ QSOs.  We use $H_o$ =
65 km s$^{-1}$ Mpc$^{-1}$, $\Omega_M = 0.3$ and $\Omega_\Lambda =
0.7$.

\section{The sources}

By selecting for very red, point-like optical sources, McMahon (1991) 
has identified a large sample of $z \ge 4$ QSOs from the Automatic
Plate Measuring survey  (Irwin, McMahon, \&\ Hazard 1991), including
the  two sources BRI 1202--0725 at $z = 4.7$ 
and BRI 1335--0415 at $z = 4.4$.
Optical spectra of both of these sources show strong associated
Ly$\alpha$ absorption (Storrie-Lombardi et al. 1996).

\subsection{1202--0725}

Omont et al. (1996) and Ohta et al. (1996) detect high order CO
line emission from 1202--0725, as well as thermal continuum
emission from warm dust at 1.35mm. The continuum and line emission
consist of a double source with an angular separation of 4$''$.
The total flux density of the source at 1.2mm is $12 \pm 3$ mJy, 
implying a far IR luminosity of $L_{\rm FIR}  = 4.2 \times
10^{13}$ L$_\odot$, assuming a spectral energy distribution (SED) typical
for an ultra-luminous infrared galaxy (ULIRG; $L_{\rm FIR} \sim
10^{12}$), where $L_{\rm FIR}$ is defined as in Helou et al. (1988)
and  Condon (1992), ie. the integrated luminosity between rest frame
wavelengths of 42 to 122 $\mu$m.
The southern source in 1202--0725 comprises about 65$\%$ of the
total.   Observations of this source with ISO at mid- to far-IR
wavelengths constrain the dust temperature to be $\simeq 70$ K
for a dust emissivity index $\beta = 1.5$ (Leech, Metcalfe, \&
Altieri 2001).  

The CO observations of Omont et al. (1996a) and Guelin et al. (2001)
indicate a difference in the CO line widths for the northern and
southern components. Gaussian fitting to the CO(5-4) line emission
profiles by Omont et al. (1996) results in $z = 4.6947$, line Full
Width at Half Maximum (FWHM) = 190 km s$^{-1}$, and an integrated line
flux density = 1.1$\pm$0.2 Jy km s$^{-1}$ for the southern source, and
$z = 4.6916$, FWHM = 350 km s$^{-1}$, and an integrated line flux
density = $1.3 \pm 0.3$ Jy km s$^{-1}$ for the northern source. This
difference is important, since it argues against gravitational lensing
as the origin for the observed double source and large apparent
luminosity.  However, Guilloteau (2001) has recently called the
difference in CO profiles into question due to the possible
contamination of the line emission from the northern source by
continuum emission at 3mm. The data presented herein support the
original conclusions of Omont et al. (1996a) and Guelin et al. (2001)
for different line profiles for the northern and southern sources in
1202--0725. 

BRI1202--0725 is a radio continuum source with a
total flux density of $315\pm80$ $\mu$Jy at 1.4 GHz
(see section 4.1),
and 141$\pm$15 $\mu$Jy at 4.9 GHz (Yun et al. 2000). 
Extrapolating the centimeter spectrum to millimeter wavelengths 
assuming a continuous powerlaw spectrum 
implies a non-thermal  contribution to the integrated 
emission at 40 GHz of 80$\mu$Jy. Conversely, extrapolating
the (sub)mm dust emission spectrum downward to 40 GHz 
implies  a thermal dust contribution of 25$\mu$Jy, assuming
$\beta = 1.5$. 

Optical imaging of 1202--0725 shows a point-like
QSO with  M$_B$ = --28.5, 
with faint near-IR and Ly$\alpha$ emission extending about
2.4$''$ north of the QSO (Hu, McMahon, \& Egami 1996). 
The optical position of the QSO given by Hu et al. 
is 0.5$''$ north of the position of the southern 
mm/cm/CO component. In the following we assume that
this offset indicates the accuracy of the
relative astrometry of the radio and optical  images.

The (3$\sigma$) limit to the Ly $\alpha$ flux from 
the northern mm/cm/CO component in 1202--0725
is $\sim 1\times10^{-17}$ erg cm$^{-2}$  s$^{-1}$
(Hu et al. 196), corresponding to a limit to the Ly $\alpha$ luminosity
of  $2\times 10^{42}$ erg s$^{-1}$. For comparison, the Ly$\alpha$
luminosities of UV dropout galaxies at $z \sim 3.09$ 
are typically $\rm 2~ to ~4 \times 10^{42}$  erg s$^{-1}$
(Steidel et al. 2000). The line emitting object detected
2.4$''$ northwest of the QSO has a Ly$\alpha$ luminosity of 
$4\times 10^{43}$ erg s$^{-1}$ (Fontana et al. 1998;
Petitjean et al. 1996). If this line emission is
powered by star formation, 
then the star formation rate is $> 20$ M$_\odot$
year$^{-1}$, depending on the amount of dust extinction 
(Fontana et al. 2000). It is also possible that this line emission is
powered by UV radiation from the QSO itself (Petitjean et
al. 1996). The lack of CO and 
thermal dust emission from this position argues for the latter, 
while the lack of N V and C IV emission lines 
argues for the former (Fontana et al. 1998; Ohta et al. 2000).


\subsection{1335--0417}

Guilloteau et al. (1997) have detected CO (5--4) line emission
from BRI~1335--0417 at $z = 4.4074$, with a
line FWHM = $420$ km s$^{-1}$ and an 
integrated line intensity of 2.8$\pm$0.3 Jy km s$^{-1}$.
The flux density of the dust continuum emission measured at 1.35mm
with the Plateau de Bure Interferometer (PdBI) at 2$''$ resolution is 
5.6$\pm$1.1 mJy, while that measured at 1.25mm with the
IRAM 30m telescope (10.6$''$ resolution) is $10.3\pm1.4$ mJy.
The implied  far IR luminosity based on the
1.25 mm measurement is $3.1 \times 10^{13}$ L$_\odot$.
The 1.35mm PdBI observations shows marginal evidence for an  extended
source, with a formal Gaussian size of $1.0 \pm 0.4''$, with major axis
oriented roughly north-south. 

No high resolution optical images have been published of
1335--0417. The DSS shows an unresolved source with M$_B$ = --27.3,
although with a pixel scale of just 1$''$ the details of the source
structure remain unknown.  BRI~1335$-$0417 has been detected
in the  radio continuum, with an integrated flux density of $220\pm
43$ $\mu$Jy at 1.4 GHz and 
$76\pm 11$ $\mu$Jy at 4.9 GHz (Carilli et al. 1999).

\section{Observations}

Table 1 lists the observing parameters for BRI~1202--0725 and
BRI~1335$-$0417. We also present a re-analysis of the D
configuration observations from  Carilli et al. (1999).

Standard amplitude and phase calibration were applied, correcting for
atmospheric opacity at high frequency, and the absolute flux density
scale was set by observing 3C~286.  Fast switching phase calibration
was employed for the high frequency observations (Carilli \& Holdaway
1999).  The 43 GHz observations were dynamically scheduled, and took
place at night under excellent weather conditions. RMS phase
variations after calibration were $\le$ 20$^\circ$. The phase
coherence  was checked by imaging a calibrator with the same
calibration cycle as that 
used for the target sources. At all times the coherence was found to
be higher than 85$\%$. 

A severe limitation at the VLA for observing broad lines at high
frequencies is the maximum correlator bandwidth of 50 MHz, and the
limited number of spectral channels (7) when using this bandwidth and
dual polarization. The bandwidth of 50 MHz corresponds to a velocity
coverage of only 350 km s$^{-1}$ at 43 GHz.  These correlator
limitations preclude a meaningful determination of the line profile,
so we chose to observe in continuum mode with two IFs (Intermediate
Frequency) with two polarizations for each IF in order to maximize
sensitivity to the 
integrated line emission (Carilli et al. 1999).  Our analysis will
necessarily assume the line widths as given by the higher order
transitions.

For the CO line observations of 1202--0725 IF1 was centered on the
emission line from the southern source, while  IF2
was off-set from this by 50 MHz. Based on the (admittedly noisy)
spectra in Omont et al. (1996a), IF1 should contain most of the
emission from the southern component, while the sum of the IFs cover
the emission from  the northern component.  The continuum for
1202--0725 was  investigated with an observation at 43 GHz. 
For 1335--0417 IF1 was centered on the emission line, 
while IF2 was centered 1400 km s$^{-1}$ off the
line.  Taking into account the limited
width of the continuum band employed, we estimate that we are missing
about 30$\%$ of the velocity-integrated line emission in 1335$-$0417.

\section{Results}

\subsection{BRI 1202-0725}

The VLA D array observations of the CO(2-1) emission from 1202--0725
are shown in Figure 1, with a spatial resolution of about 2$''$.
Figure 1a shows the summed emission in IF's 1 and 2 for
the 40 GHz observations. Shown in greyscale on
Figure 1a is the Ly $\alpha$ image of 1202--0725 from Hu et
al. (1996). Again, we have aligned the position of the optical
QSO with the peak in the radio continuum, CO, and thermal dust
emission for the southern source.
Figures 1b and 1c show IFs 1 and 2, respectively, for the 40 GHz
observations. Figure 1d shows the continuum image at 43 GHz.

Table 2 lists the results for the two components in 1202--0725.
Column 3 lists the position of the measured  CO components.
Column 4 lists the 7mm continuum flux densities at these
positions, column 5 lists the observed CO flux densities, and 
column 6 lists the velocity integrated CO emission. 
Column 7 lists the 1.4 GHz continuum flux densities, while column 8
lists the CO luminosities (in K km s$^{-1}$ pc$^{2}$). 

The southern CO component is clearly detected in IF1 of the 40 GHz
observations (Fig. 1b), but is not seen in IF2, nor in the 43 GHz
continuum image (Figs. 1c and 1d).
The northern component shows up
in both IF channels for the 40 GHz observations (Figs 1a,b,c), 
but is not seen in the 43 GHz continuum image (Fig 1d).
From the 43 GHz image  we set a 2$\sigma$ upper limit of 0.16 mJy for
the  43 GHz continuum emission from both the northern and southern
components in 1202--0725. 

The lack of continuum emission at 43 GHz
from  the northern component in 1202--0725 is significant in regard to
the recent possible detection of $0.8\pm0.15$ mJy of
continuum emission at 3mm by Guilloteau (2001). 
The implied spectrum from 3mm to 7mm must be rising
with a powerlaw index $> 1.9$. It also implies that the 
observed emission in IFs 1 and 2 at 40 GHz for the northern
source is CO(2-1) line emission (Figs. 1a,b,c), and
hence that the velocity profiles for the CO lines from the northern 
and southern sources are different. 

Figure 2 shows the results from the B array observations with a
resolution of about $0.25''$.  The southern source in IF1 (Fig 2b)
appears as two unresolved, roughly equal components (flux
densities $\sim 0.41\pm0.12$ mJy),
separated by 0.3$''$.  The implied lower limit to the
(redshift corrected) brightness temperature for these
components is about 25 K. 
The northern source in IFs 1
and 2 (Fig 2a) is marginally detected with a 
peak surface brightness of $0.25 \pm 0.08$ mJy beam$^{-1}$.  
This implies two, or more, compact components with
flux densities below our detection threshold, or 
diffuse emission on a scale $\ge 0.5''$. 

Figure 3 shows the radio continuum image of 1202--0725 at 1.4 GHz,
with a resolution of about $2''$.
Both components are detected, with flux densities
as listed in Table 2. 
The total 1.4 GHz flux density of 1202--0725 has been measured
three times over the course of three years with values
of $240\pm40$ $\mu$Jy (Yun et al. 2000), 
$305\pm60$ $\mu$Jy, and $390\pm40$ $\mu$Jy. 
In our analysis we adopt the mean value of 
315 $\mu$Jy. It is possible that the source
is variable, although making accurate flux density
measurements at this level is difficult due to confusion
problems arising in wide-field imaging at 1.4 GHz. 

\subsection{BRI 1335--0417}

The low resolution image of the CO(2-1) emission from 1335--0417,
as reproduced from Carilli et al. (1999),
is shown in Figure 4a, along with the off-line channel in Figure
4b. For this image 
the UV-data were naturally weighted in order to maximize sensitivity,
at the expense of resolution. The rms noise on the image is 
0.11 mJy beam$^{-1}$ and the resolution is about $1.6''$. 
The CO emission appears
extended north-south in this image by about 1$''$. 

In order to investigate this extension in more detail, we re-imaged
the data using uniform weighting of the visibilities, which optimizes
resolution at the expense of sensitivity.  The result is shown in
Figure 4c, with a rms noise of 0.14 mJy and resolution of
about $1.3''$.  The source is comprised of two components 
separated by 1.3$''$.  Gaussian fitting to each component
shows that they are unresolved, with upper limits to their sizes of
about 1.1$''$. Table 2 lists the component positions and flux densities.

The position of the 1.35mm continuum peak is located within
0.2$''$ of the southern CO component, and the mm continuum source also
shows marginal evidence for a north-south extension on the scale of
1$''$ (Guilloteau et al. 1997).  The
optical QSO position is within 0.1$''$ of the southern 
source position.

Figure 5 shows the high resolution (B array) image of the CO(2-1)
emission from 1335--0417. The naturally weighted beam is roughly
circular with FWHM = $0.17''$. Nothing is detected in this high
resolution image to 
a 2$\sigma$ surface brightness limit of 0.18 mJy beam$^{-1}$.
Non-detection at high resolution could mean that the emission is
diffuse on scales larger than the resolution with a (redshift
corrected) brightness temperature $\le 22$ K.  On the other hand, a
number of compact components distributed over 1$''$ with higher
brightness temperature is certainly not precluded, eg. the data allow
for four small ($\le 0.17''$) components each with brightness
temperatures $\ge$ 22 K.

\section{Analysis}

\subsection{Masses}

The H$_2$ gas masses can be calculated from the values of $L'$ in Table
1 assuming a value of   X = the H$_2$ mass-to-CO(1-0)
luminosity conversion factor in M$_\odot$ (K km s$^{-1}$ pc$^2$)$^{-1}$.
A value of $\rm X \simeq 4.6$  
is applicable to Galactic  Giant Molecular Clouds (Dame et
al. 1987;  Strong et al. 1988; Bronfman et al. 1988), while for ULIRGs
Downes and Solomon (1998)
find a value of $\rm  X \simeq 0.8$. This calculation also requires an
extrapolation from the CO(2-1) measurements to the CO(1-0)
luminosity. Assuming constant brightness temperature and
using a value of X appropriate to ULIRGs
leads to molecular gas masses of about $5\times10^{10}$ M$_\odot$
for the CO emitting components in 1202--0725 and 1335--0417. 

We can also calculate the gravitational masses using the observed
sizes and line widths for the sources, and assuming 
a disk of gas in Keplerian rotation.  
For 1202--0725 south we use a line width of 190 km s$^{-1}$
and a radius corresponding to half the separation of the
two components = 0.15$''$ = 1.1 kpc, leading to an
enclosed mass of $0.23 \times 10^{10}$ sin$^{-2}(i)$ M$_\odot$.
For 1335--0417 we do not have spatially resolved spectroscopy,
so we adopt a line width of 420 km s$^{-1}$ and a radius 
of half the separation of the two components = 0.65$''$ = 4.7 kpc,
leading to an enclosed mass of $4.8 \times 10^{10}$ sin$^{-2}(i)$
M$_\odot$. 

An upper limit to $i$ can be derived by assuming that the 
molecular gas mass dominates the total enclosed mass. 
Using $\rm X = 0.8$ for 1202--0725 south we find
$i \le 12^o$, while for 1335--0417 we find $i \le 43^o$.
These small angles lead to a significant problem in
that the implied rotational velocities are then extremely
large,  $\ge$ 400 km s$^{-1}$. 
To obtain rotational velocities of $\sim 250$ km s$^{-1}$,
more typical of large spiral galaxies, would require a 
further reduction in the conversion factor to
$\rm X \le 0.2$.  However, such a low X value would violate the 
minimum mass conditions dictated by optically thin CO emission,
for which $\rm X \sim 0.35$ for an excitation temperature of 50 K
and assuming a Galactic CO abundance (Solomon et al. 1997). 
It may be that these systems are extremely massive, 
or perhaps that  the apparent CO luminosities are magnified by 
gravitational lensing (see section 6).  

\subsection{Continuum-to-Line Ratios}

We now consider the continuum-to-line
ratio: ${{L_{\rm FIR}}\over{L'_{\rm CO(1-0)}}}$.
A non-linear relationship between this ratio
and $L_{\rm FIR}$ has been found by Solomon et al. (1997),
in which  values of 5 to  50 have been found
for Galactic Giant Molecular Clouds and for nearby galaxies with
$L_{\rm FIR} \le 10^{10}$, while values of 80 to 250 have been
found for ULIRGs.   

For the QSOs discussed herein we derive the 
$L'$(CO(1-0)) from $L'$(CO(2-1))
assuming the same brightness temperature for the two transitions. 
Considering the integrated 
properties of 1202--0725  leads to: 
${{L_{\rm FIR}}\over{L'_{\rm CO(1-0)}}} 
= {{4.2\times10^{13} \rm L_\odot}\over{1.2\times 10^{11} \rm K~ km~ s^{-1}~
pc^{2}}} = 350$. For 1335-0417 we find:
${{L_{\rm FIR}}\over{L'_{\rm CO(1-0)}}} 
= {{3.1\times10^{13} \rm L_\odot}\over{9.6\times 10^{10} \rm K~ km~ s^{-1}~
pc^{2}}} = 323$. 

\subsection{CO excitation conditions}

Figure 6 shows the CO ladder for the various transitions
observed in 1202--0725 and 1335--0417. 
The data have all been normalized to the velocity integrated
line flux for CO(5-4). For comparison, we have
included the CO ladder observed for the best
studied nuclear starburst M82 (G\"usten et al. 1993;
Mao et al. 2000),
and the CO ladder for the integrated emission from the
the Milky Way disk inside the solar radius (excluding the Galactic
center) as seen by COBE (Fixsen, Bennett, \& Mather 1999). The
excitation conditions for the two QSOs  follow those seen in M82,
with roughly constant brightness temperature (intensity $\propto$ 
frequency$^2$) up to CO(5-4), and then  roll-off to higher order
transitions. This behavior is very different than that seen
for the disk of the Milky Way, for which the integrated line
flux peaks at the CO(3-2) line. 

We have used a standard one component LVG model simulating a
spherical cloud to interpret the observed line ratios
(see also Ohta et al. 1998). Input
parameters are the kinetic temperature, the H$_2$
density, and the CO column density, the latter in the form 
$\rm N(CO) = 3.08 \times 10^{18} \times n_{CO} \times
[{{\Delta V}\over{grad V}}]$ cm$^{-2}$, with grad $V$ denoting
the velocity gradient in km s$^{-1}$ pc$^{-1}$, n(CO) being the
nominal CO number density in cm$^{-3}$, and $\rm \Delta V$ 
describing  the FWHM in km s$^{-1}$. Given the inhomogeneity of the
interstellar medium in the Milky Way and nearby galaxies, 
our assumption of uniform physical conditions is crude but 
appropriate for a source with four detected lines from the 
main CO species and no information on rare isotopomers. The 
cosmic background was assumed to be at a temperature of 15 K.

Due to the lack of constraints, 
there is a clear degeneracy in LVG modeling 
between various parameters such as 
the temperature and  density (G\"usten et al. 1993). 
Our modeling is not meant as an exhaustive analysis, but
is merely representative of the types of conditions that
can give rise to the observed line ratios. Adopting
a kinetic temperature of 70 K, corresponding to the temperature
derived from the spectrum of thermal dust emission (Leech et al. 2001),
we find that the data can be reasonably fit with
a model in which: N(CO) = $5 \times 10^{19}$ cm$^{-2}$,
n(H$_2$) = $2\times 10^4$ cm$^{-3}$,
and grad $V$ = 1 km s$^{-1}$ pc$^{-1}$. 
Raising the temperature to 100 K decreases the
required density by a factor two. A more 
thorough analysis of the excitation conditions
in 1202--0725, including new, spatially resolving observations of the
high order transitions will be given elsewhere (Kohno et al. 2001, in
preparation). 

In the likely case that photon dominated regions (PDR) provide a more
realistic approach, the density must be larger than the `critical'
density to efficiently excite the mid- and high-J CO lines and to
avoid subthermal excitation (e.g. Koester et al. 1994).
With line temperatures being highest for the CO J = 2-1 or 3-2
transitions, critical densities, at which collisional deexcitation
rates exceed radiative decay rates, are $10^4$ to $10^5$ cm$^{-3}$.

\section{Discussion}

We briefly compare the properties of 1202--0725 and 1335--0417 
to those found in nuclear starburst galaxies seen at lower redshift.
Yun et al. (2000) found that the ratio of radio-to-far IR luminosity
for both these sources is within the range defined for 
active star forming galaxies based on the tight radio-to-far IR correlation
(Condon 1992), although in both cases the ratio falls at 
the high end of the normal range, suggesting a possible contribution
to the radio emission from the AGN. 
For 1202--0725 there is marginal evidence that the radio emission 
is time variable. Variability would rule-out a starburst origin for
the radio emission. Further radio continuum monitoring is in progress
to test this interesting possibility. 

The measured (redshift corrected) brightness temperature for the
CO(2-1) emission from the compact components in the southern
source in 1202--0725 is $\ge 25$ K.
This limit is  consistent with values  seen for the CO(2-1) 
emission from starburst nuclei
of ULIRGs, for which brightness temperatures 
of 30 K to 60 K have been measured (Downes and Solomon 1998).

The excitation conditions for the CO for these two sources follow
roughly those seen for the nuclear starburst galaxy M82, but are very
different than those expected for the disk of a normal spiral galaxy.  
Assuming a Galactic abundance for CO of 
$\rm {{[CO]}\over{[H_2]}} = 5\times 10^{-5}$ (eg. Wilson et al. 1986)
implies an 
H$_2$ column density  of order 10$^{24}$ cm$^{-2}$. 
This value  is similar to
the molecular gas column density seen toward ULIRGs (Downes \& Solomon
1998),  but is an order
of magnitude larger than that observed in the lower luminosity 
nuclear starburst galaxies M82  and NGC 253 (Harrison, 
Henkel, \& Russell 1999, Mao et al. 2000). A 
low metallicity would suppress CO emission and would yield even higher
H$_2$ column densities. It would be very interesting to measure the 
CO(1-0) line of the source.  A direct comparison of CO(1-0) and 2-1 line
intensities would provide significant information on optical depths,
since the 1-0 line may be optically thin, while the 2-1 line is likely
optically thick.

We have found that the continuum-to-line ratio,
${{L_{\rm FIR}}\over{L'_{\rm CO(1-0)}}}$,
is about 335 for both sources. This ratio is at the high end
of those seen for ULIRGs, and suggests a continuation of
the non-linear trend for increasing continuum-to-line ratios with
increasing far IR luminosity. 
Considering $L_{\rm FIR}$ to be a measure of star
formation rate, and $L'$(CO(1-0)) to be a measure of molecular gas
mass, it has been suggested that this non-linear relation might
imply a higher star formation efficiency ($\equiv \rm 
{{Star~Formation~Rate}\over{Gas~Mass}}$) in higher luminosity
galaxies, in particular for galaxies with $L_{\rm FIR} \ge 10^{11}$
L$_\odot$ (Solomon et al. 1997). 
For dense nuclear  starbursts
a number of groups (Solomon et al. 1997; Mao et
al. 2001; Weiss et al. 2001) have shown that the densities
are such that the entire interstellar medium in the starburst regions
may be molecular, and that the CO(1-0) emission may be dominated
by this molecular inter-cloud medium, as opposed to being
from the denser star forming clouds themselves. 
This phenomenon would contribute to the non-linear relationship
between $L_{\rm FIR}$ and $L'$(CO(1-0)).
For 1202--0725 and 1335-0417 there
is also the obvious possibility of dust heating by the AGN,
in which case the continuum-to-line ratio cannot be 
interpreted in the context of star formation.

Downes and Solomon (1998) show that for the nuclear starburst
regions in ULIRGs the H$_2$ mass-to-CO(1-0) luminosity conversion factor,
X, is a factor four or so below the Galactic disk value.
Even for this low value of X, we find that the CO emitting regions in
1202--0725 and 1335--0417 must be
close to face-on in order to avoid having the gas mass 
exceed the gravitational mass, implying perhaps unreasonably
large rotational velocities. While this problem is mitigated
somewhat by lowering X even further, the required X values become
comparable to, or lower than,  the  minimum values dictated by
optically thin CO emission (Solomon et al. 1997). 

One way of circumventing this mass problem would be to assume that
the source is magnified by strong gravitational lensing. 
Magnification by  a factor three or so would avoid unphysical 
values of X.  The very large IR and CO luminosities, and the
double structure of the sources in the CO line
emission, and in the non-thermal radio continuum and thermal mm
continuum for 1202--0725, could possibly indicate gravitational 
lensing.  On the other hand, neither source is double at
optical wavelengths, arguing against lensing,  although the
possibility of differential obscuration along the two lines-of-sight  
complicates this conclusion (Hu et al. 1996). More telling
is the difference in CO(2-1) line profiles, which is difficult,
although perhaps not impossible, 
to explain in the context of gravitational lensing. 
Also, double sources arise from strong gravitational lensing
of very compact emitting regions (sizes $\le$ few pc).
For extended emitting regions ($\ge 100$ pc), such as must 
be the case for the thermal CO and mm continuum emission,
strong lensing will only occur if the extended emission regions
cross a caustic in the source plane (Blandford \&
Narayan 1992). Such a phenomenon usually leads
to more complex geometries, like arcs or rings, as in APM 0827+525
and IRAS 10214 (Lewis et al. 2001; Scoville et al. 1995). 
More sensitive, high resolution  imaging at cm and mm wavelengths is
required  to address this interesting question. 

Overall, the physical conditions in the molecular gas and
dust in these systems are similar to those observed
in  nuclear starbursts at low redshift, including: 
(i) the radio-to-far IR luminosity ratio, (ii) the CO
brightness temperature, (iii) the CO excitation conditions,
(iv) the CO column densities,  and 
(v)  the CO line-to-dust continuum ratio.

\vskip 0.2truein 

The National Radio Astronomy Observatory (NRAO) is operated
by Associated Universities, Inc. under a cooperative agreement with the
National Science Foundation. We thank E. Hu for allowing us to
reproduce the Ly$\alpha$ image, S. Myers for discussions concerning
gravitational lensing, and the referee for many important comments. 


\clearpage
\newpage

\centerline{\bf Figure Captions}

F{\scriptsize IG}. 1a.--- The contours
show the VLA image of CO(2-1) emission from 1202--0725 
at $z = 4.691$.  The spatial resolution (FWHM) is $2.6'' \times 1.8''$
with the major axis position angle = 0$^o$, and the rms noise
on the image  is $\sigma = 0.07$ mJy beam$^{-1}$. 
This frame shows the sum of the two IFs with a mean
frequency of 40.510 GHz and a total bandwidth of 100 MHz. 
The contour levels are: -0.13, 0.13, 0.26, 0.39 mJy beam$^{-1}$. 
The greyscale shows the narrow band image of 
the Ly$\alpha$ emission (Hu et al. 1996). The crosses
in this  and subsequent images of 1202--0725, mark
the two peak positions of the CO emission in this image. \\
1b. --- The image of IF1 (40.485 GHz) and $\sigma = 0.1$ mJy
beam$^{-1}$.
The contour levels are: -0.19, 0.19, 0.38, 0.57, 0.76 mJy
beam$^{-1}$. \\
1c. --- The image of IF2 (40.535 GHz) with the same contours as 1b and
$\sigma = 0.1$ mJy beam$^{-1}$.  \\ 
1d. --- The 43 GHz continuum image of 1202--0725 with a resolution of
$3'' \times 1.6''$, with the major axis position angle = $-20^o$, and
$\sigma = 0.08$ mJy.   
The contour levels are: -0.2, -0.1, 0.1, 0.2, 0.3 mJy beam$^{-1}$. 

F{\scriptsize IG}. 2a.--- The VLA image of the CO(2-1) emission from
the northern component in 1202--0725 using the sum of IFs 1 and 2 at
40 GHz at a resolution of $0.30'' \times 0.21''$ with the major axis
position angle = 11$^o$, and $\sigma = 0.08$ mJy beam$^{-1}$.  The
contour levels are:  -0.2, -0.1, 0.1, 0.2, 0.3 mJy beam$^{-1}$. \\
2b.--- The VLA image of the CO(2-1) emission from the
southern component in 1202--0725  for IF1 at 40.485 GHz,
and $\sigma = 0.12$ mJy beam$^{-1}$.  
The contour levels and resolution are the same as Figure 2a. \\

F{\scriptsize IG}. 3.--- An image of the radio continuum emission
from 1202--0725 at 1.4 GHz made with the VLA at a resolution of
$2.0'' \times 1.5''$, major axis position angle = -30$^o$,
and $\sigma = 16$ $\mu$Jy beam$^{-1}$.  
The contour levels are a geometric progression in the square
root of two with the first level being 35$\mu$Jy beam$^{-1}$. 

F{\scriptsize IG}. 4a.--- The contours
show the VLA image of CO(2-1) emission from 1335--0417
at $z = 4.4074$.  The spatial resolution is  $1.4'' \times
1.3''$ with major axis PA = 39$^o$,
and $\sigma = 0.11$ mJy beam$^{-1}$.  
This frame shows the first IF at 42.635 GHz 
and a bandwidth of 50 MHz. 
The contour levels are: -0.24, -0.12, 0.12, 0.24, 0.36, 0.48, 0.50 mJy
beam$^{-1}$.  The crosses in this and subsequent images, 
mark the positions of the two peaks as seen in Figure 4c. \\
4b. --- The same as 4a but for the second IF at 42.835 GHz. \\
4c. --- The same as 4a but using uniform weighting of the 
visibilities, leading to a spatial resolution of
$1.4'' \times 1.3''$ with major axis PA = 39$^o$,
and $\sigma = 0.14$ mJy beam$^{-1}$.  
The contour levels are: -0.26, -0.13, 0.13, 0.26, 0.39, 0.52 mJy
beam$^{-1}$.

F{\scriptsize IG}. 5. --- An image of the CO(2-1) emission from 1335--0417
at a spatial resolution of $0.17''$, and $\sigma = 0.09$ mJy
beam$^{-1}$.   This frame shows the first IF at
42.635 GHz  and a bandwidth of 50 MHz. 
The contour levels are: -0.28, -0.14, 0.14, 0.28, 0.42 mJy
beam$^{-1}$.

F{\scriptsize IG}. 6. --- The CO ladder for the two high redshift QSOs
in this paper and using data from Omont et al. (1996), Ohta et
al. (1996), and Guilloteau et al. (1997) for the other
transitions.  The open squares are the results for BRI 1202--0725.
The open triangles are those for 1335--0417. The solid triangles are
the data for the CO ladder for the integrated emission from the
the Milky Way disk inside the solar radius (excluding the Galactic
center) as seen by COBE (Fixsen, Bennett, \& Mather 1999).
The open circles are the results for the
starburst nucleus of M82 (G\"usten et al. 1993; Mao et al. 2000).
The ordinate is the velocity integrated line
flux density, normalized to the CO(5-4) line, except for the Milky
Way, for which the values are normalized at CO(3-2). The long dash
line shows the line strengths increasing as frequency squared. The
short dash line shows an LVG model with T$_{\rm kin}$ = 100 K and 
n(H$_2$) = $1\times 10^4$ cm$^{-3}$,
and the dotted line shows the LVG model with  T$_{\rm kin}$ = 70 K and
n(H$_2$) = $2\times 10^4$ cm$^{-3}$.

\clearpage
\newpage

\begin{deluxetable}{cccc}
\tablewidth{0pc}
\tablecaption{VLA Observations} 
\tabletypesize{\small}
\tablehead{
Source & Date & Configuration & Frequency   \\
~ & ~ & ~ & GHz \\
}
 \startdata
1202--0725  & March 1999 & D (1km) & 40.485, 40.535 \\
1202--0725  & May 2001 &  B (10km) &   40.485, 40.535 \\
1202--0725  & November 2001 & D & 43.315, 43.365 \\
1202--0725  & August 1999 &  A (30km) & 1.365, 1.415  \\
1335--0417 & December 1999 & B & 42.635, 42.835 \\
1335--0417 & March 1999 & D &  42.635, 42.835 \\
\tableline
\enddata
\end{deluxetable}

\clearpage
\newpage

\begin{deluxetable}{ccccccccc}
\tablewidth{0pc}
\tablecaption{Observed Parameters} 
\tabletypesize{\tiny}
\tablehead{
Source & $z$ & Position & S(7mm) &  S(CO(2-1)) & S
$\Delta$V & S(1.4 GHz) & L'(CO(2-1)) \\
~ & ~  &  J2000 & mJy & mJy & Jy km s$^{-1}$ & mJy &
$\times10^{10}$ K km s$^{-1}$ pc$^{2}$ \\
}
\startdata
1202--0725 South & 4.695 & 12~05~23.12~--07~42~32.9$\pm0.3''$ & 
$0.16\pm0.08$ & $0.77\pm0.10$ & $0.23\pm0.04$ & $0.16\pm0.016$ &
$5.5 \pm 1.0$ \\
1202--0725 North & 4.692 & 12~05~22.98~--07~42~29.9$\pm0.3''$ & 
$-0.13\pm0.08$ & $0.44\pm0.07$  & $0.26\pm0.05$ &  $0.23\pm0.016$ 
& $6.2\pm1.3$  \\
1335--0417 South & 4.407 & 13~38~03.40~--04~32~35.4$\pm0.3''$ & 
$-0.03\pm0.12$ & $0.67\pm0.14$ & $0.26\pm0.06$  & $0.22\pm0.043$ &
$5.7 \pm 1.1$  \\
1335--0417 North & 4.407 & 13~38~03.42~--04~32~34.1$\pm0.3''$ & 
$0.07\pm0.12$  & $0.45\pm0.14$ &  $0.18\pm0.06$ & -- &
$3.9 \pm 1.1$ \\
\tableline
\enddata
\end{deluxetable}

\clearpage
\newpage

\begin{figure}
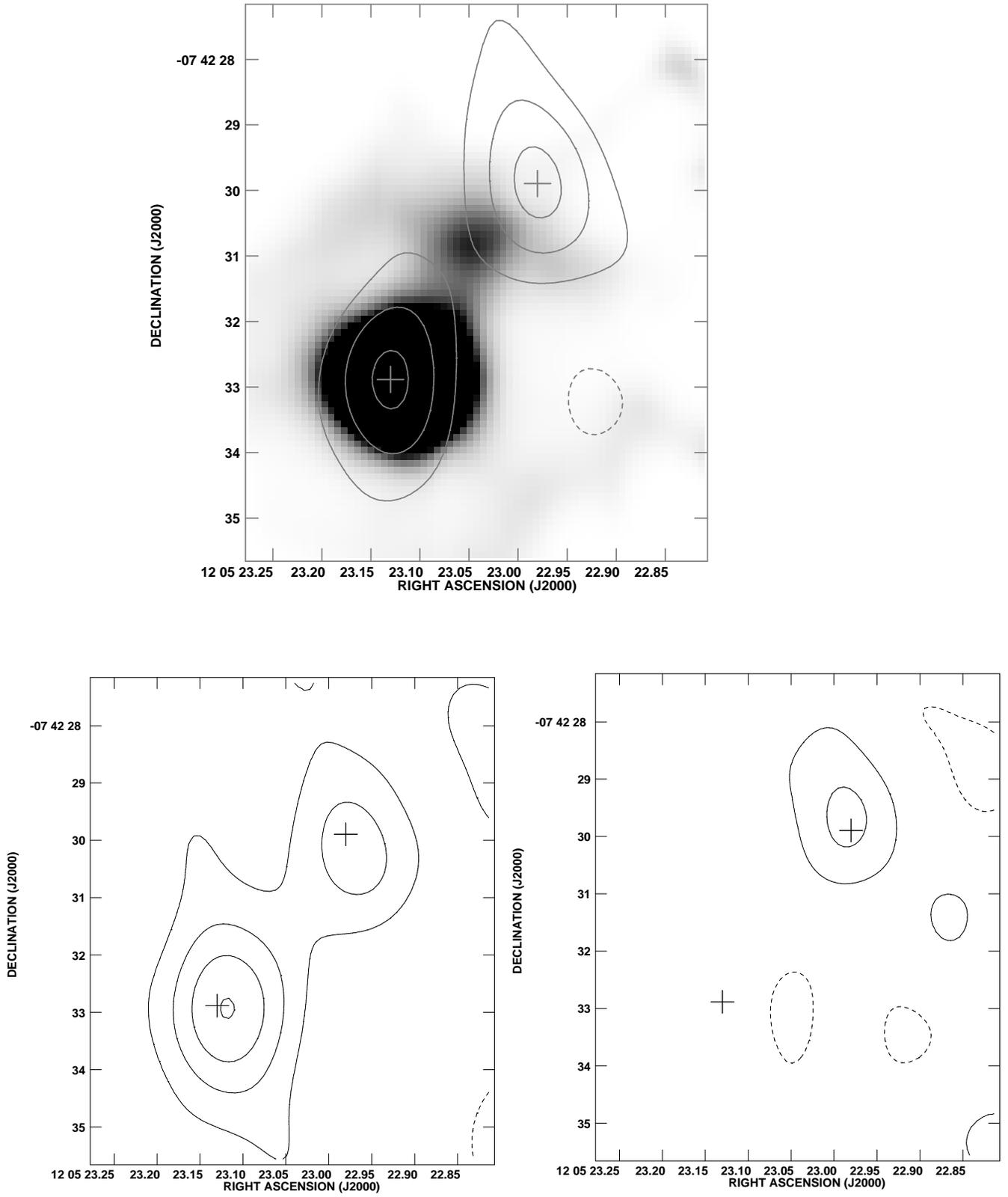

\vskip -0.4in
\hskip 1in
\psfig{figure=ccfig1a.ps,width=4in}
\psfig{figure=ccfig1b.ps,width=3.5in}
\vskip -4.1in
\hspace*{3.5in}
\psfig{figure=ccfig1c.ps,width=3.5in}
\caption{1a -- upper; 1b -- lower left; 1c -- lower right; 1d -- next page}
\end{figure}

\begin{figure}
\psfig{figure=ccfig1d.ps,width=5in}
\end{figure}

\begin{figure}
\psfig{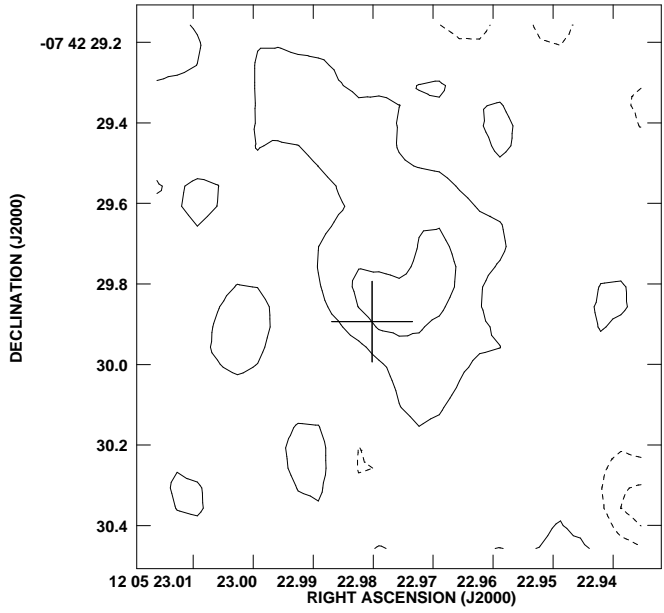}
\psfig{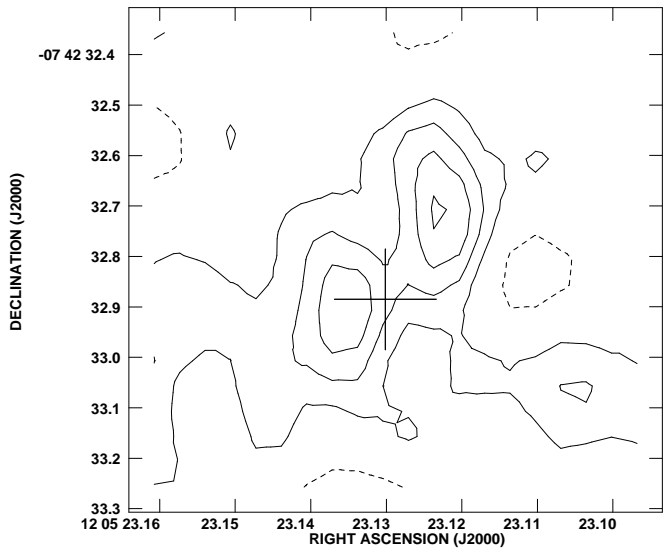}
\caption{2a -- upper; 2b -- lower}
\end{figure}

\begin{figure}
\psfig{figure=ccfig3.ps,width=5in}
\caption{}
\end{figure}

\begin{figure}
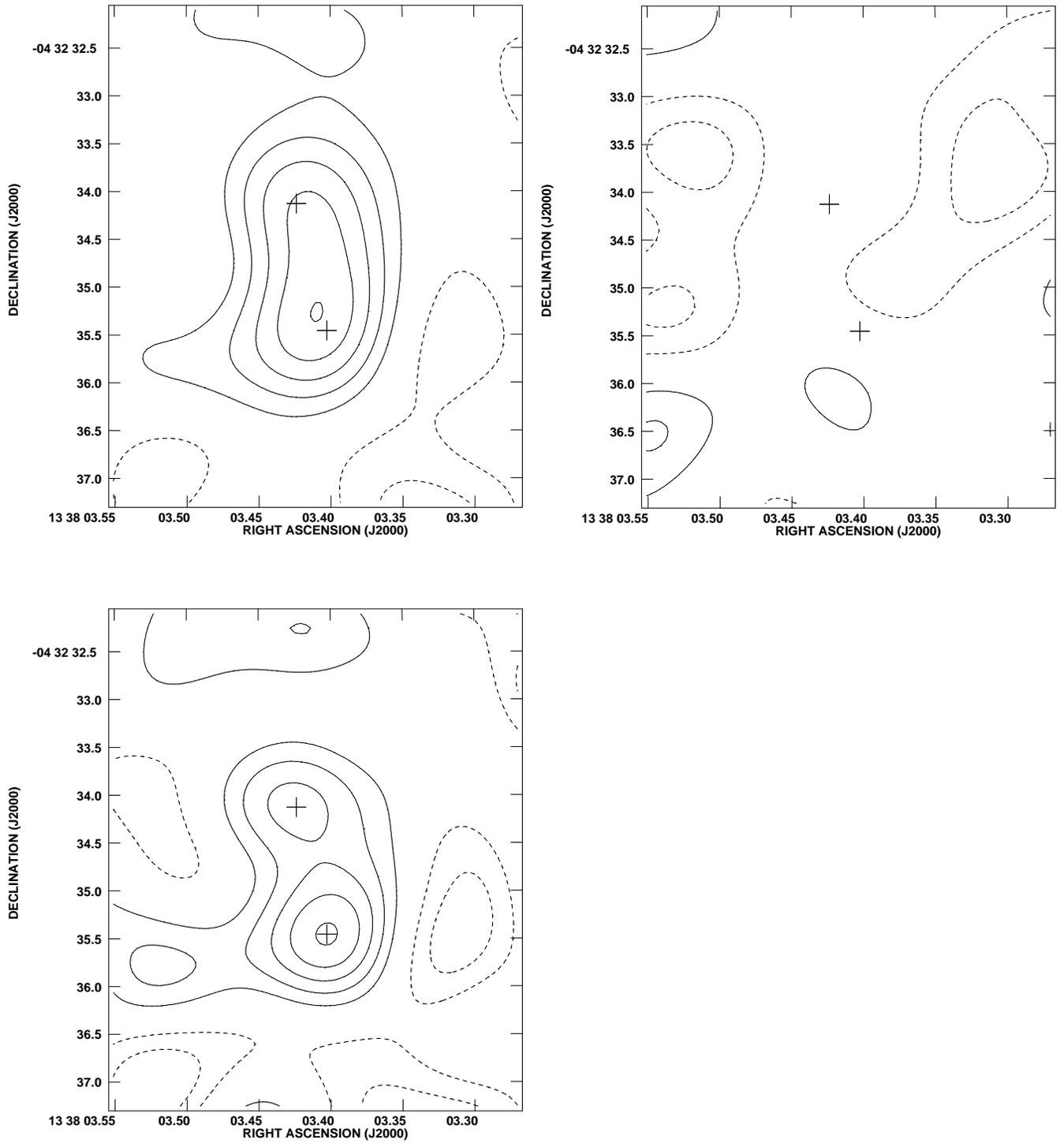

\vskip -0.4in
\psfig{figure=ccfig4a.ps,width=3.5in}
\vskip -4in
\hspace*{3.5in}
\psfig{figure=ccfig4b.ps,width=3.5in}
\psfig{figure=ccfig4c.ps,width=3.5in}
\caption{4a -- upper left; 4b -- upper right; 4c -- lower}
\end{figure}

\begin{figure}
\psfig{figure=ccfig5.ps,width=5in}
\caption{}
\end{figure}

\begin{figure}
\psfig{figure=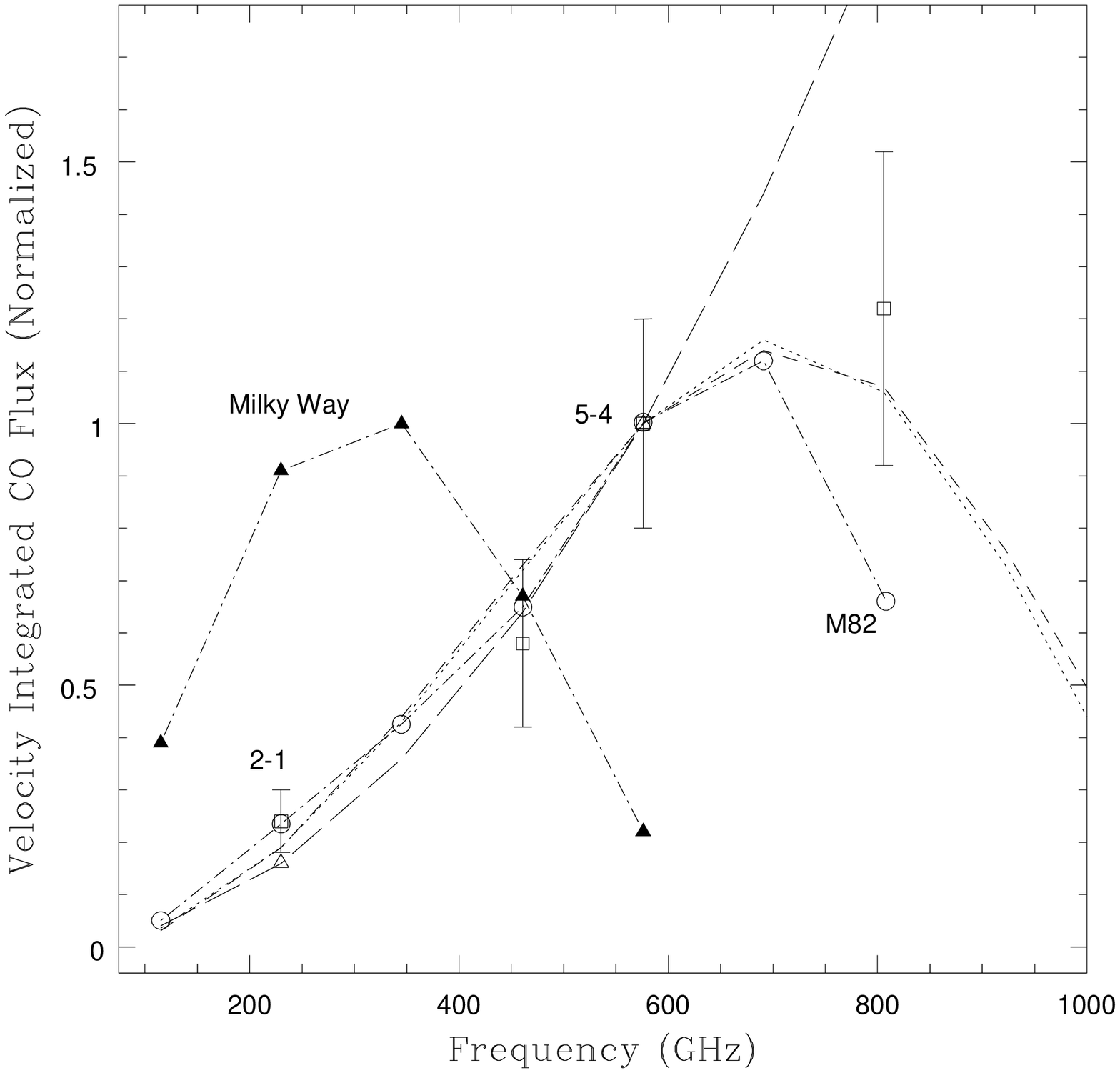,width=6in}
\caption{}
\end{figure}

\clearpage
\newpage


\begin{references}

\reference{}Blain, A.W. Jameson, A., Smail, I., Longair, M.S., Kneib,
 J.-P., \& Ivison, R.J. 1999, MNRAS, 309, 715

\reference{}Blandford, R.D. \& Narayan, R. 1992, ARAA, 30, 311

\reference{} Bronfman, L., Cohen, R.S., Alvarez, H., May, J., 
\& Thaddeus, P. 1988, ApJ, 331, 181

\reference{}Carilli, C. L. \& Yun,  M. S. 1999, ApJ, 513, L13

\reference{}Carilli, C. L. \& Holdaway, 
M. A. 1999, Radio Science, 34, 817

\reference{} Carilli, C.L.,  Bertoldi, F., Rupen, M.P. et al.
2001a, ApJ, 555, 625

\reference{} Carilli, C.L., Bertoldi, F., Omont, A., Cox, P., McMhaon,
 R.G., \& Isaak, K.   2001b, AJ, in press

\reference{}Carilli, C.L., Menten, K.M., \& Yun, M.S.
1999, ApJ, 521, L25

\reference{}Condon, J. J. 1992, ARAA, 30, 575

\reference{}Cox, P., Omont, A., Bertoldi, F. et al. 2001, A \& A
(letters), submitted 

\reference{}Dame, T., Ungerechts, H.,  Cohen, R. S. et al. 1987, ApJ,
322, 706 

\reference{}Downes, D., \& Solomon, P.M. 1998, ApJ, 507, 615

\reference{}Djogovski, S.G. 1999, in {\sl The Hy-redshift Universe,}
(ASP: San Francisco), eds. A.J. Bunker \& W.J.M. van Breugel, p. 397

\reference{}Fan, X. et al. 2001, AJ, 121, 54

\reference{} Ferrarese, L. \& Merritt, D. 2000, ApJ (letters),
539, 9

\reference{} Fixsen, D.J., Bennett, C.L., \& Mather, J.C. 1999,
ApJ, 526, 207

\reference{}Franceschini, A., Hasinger, G., Mayaji, T., \& Malquori,
D. 1999, MNRAS (letters), 310, 5

\reference{}Fontana, A., D'Odorico, S., Giallongo, E., Cristiani, S.,
Monnet, G., \& Petitjean, P. 1998, AJ, 115, 1225

\reference{} Gebhardt, Karl et al. 2000, ApJ (letters), 539, 13

\reference{}Guelin, M. et al. 2001, in preparation

\reference{}Guilloteau, S., Omont,A., 
McMahon, R.G., Cox,P., \& Petitjean,P. 1997, A\&A, 328, L1

\reference{}Guilloteau, S., Omont, A., Cox, P., 
McMahon, R. G., \& Petitjean, P. 1999, A\&A, 349, 363

\reference{}Guilloteau, S. 2001, in {\sl Science with Large Millimeter
Arrays}, ed. A. Wootten, (San Francisco: ASP)  in press

\reference{}G{\"u}sten, R., Serabyn,
E., Kasemann, C., Schinckel, A., Schneider, G., Schulz, A., \& Young,
K. 1993, ApJ, 402, 537 

\reference{}Harrison, A., Henkel, C., \& Russell, A. 1999, MNRAS 303,
157 

\reference{}Helou, G., Khan, I.R., Malek, L., \& Boehmer, L. 
1988, ApJ (supplement), 68, 151

\reference{}Hu, E., McMahon, R.G., \&\ Egami, E. 1996, ApJ, 459, L53

\reference{}Irwin, M., McMahon, R. G., \&
Hazard, C. 1991, in The Space Distribution of Quasars,
ed. D. Crampton, (San Francisco: PASP), p. 117 

\reference{}Kawabe, R., Kohno, K., Ohta,
K., \& Carilli, C. 1999, in Highly Redshifted Radio Lines,
eds. C. L. Carilli, S. J. E. Radford, K. M. Menten, \& G. I.
Langston, (San Francisco: PASP), p. 48

\reference{}Kauffmann, G. \& Haehnelt, M. 2000, MNRAS, 311, 576

\reference{}Koester, A., Stoerzer, H., Stutzki, J., \& Sternberg,
A. 1994, A\& A 284, 545 

\reference{}Leech, K.J., Metcalfe, L., \& Altieri, B. 2001,
MNRAS, in press, (astroph 0109033)

\reference{}Lewis, G.F, Carilli, C., Papadopoulos, P., \& Ivison, 
R.J. 2001, MNRAS (letters), in press (astroph ??)

\reference{}Mao, R.Q., Henkel, C., Schulz, A. et al. 2000, A\& A 358, 433

\reference{}McMahon, R. G. 1991, in The Space
Distribution of Quasars, ed. D. Crampton, (San Francisco: PASP),
p. 129 

\reference{}Ohta, Kouji, Yamada, T., Nakanishi,
K., Kohno, K., Akiyama, M., \& Kawabe, R. 1996, Nature, 382, 426

\reference{}Ohta, Kouji, Matsumoto, Tsuyoshi,
Maihara, Toshinori et al.  2000, PASJ, 52, 557

\reference{}Ohta, Kouji,  Nakanishi, Kouichiro,
Akiyama, Masayuki et al. 1998, PASJ, 50, 303

\reference{}Omont, A., McMahon, R. G., Cox,
P., Kreysa, E., Bergeron, J., Pajot, F., \& Storrie-Lombardi,
L.J. 1996a, A\&A, 315, 1 

\reference{}Omont, A., Petitjean, P.,
Guilloteau, S., McMahon, R. G., Solomon, P. M., \& Pecontal, E. 1996b,
Nature, 382, 428

\reference{}Omont, A., Cox, P.; Bertoldi, F.,
McMahon, R. G., Carilli, C., \& Isaak, K. G 2001, A\& A, 374, 371


\reference{}Petitjean, P., Pecontal, E., Valls-Gabaud, D. \& Charlot,
S. 1996, Nature, 380, 411

\reference{}Richstone, D., Ajhar, E. A., Bender, R. et al.
1998, Nat. Supp., 395A, 14

\reference{}Sanders, D.B., Phinney, E. S., Neugebauer, G.,
 Soifer, B. T., \& Matthews, K. 1989, ApJ, 347, 29

\reference{}Scoville, N.Z., Yun, M.S., R. L. Brown, \& P. A. Vanden
Bout 1995, ApJ (letters), 449, L109

\reference{}Solomon, P.M., Radford, S.J.E., \&\  Downes, D. 1992,
Nature, 356, 318

\reference{}Solomon, P.M., Downes, D., Radford, S., \& Barrett, 
J.W.  1997, ApJ, 478, 144

\reference{}Solomon, P.M. 2001, in {\sl Starburst Galaxies Near 
and Far}, eds. D. Lutz \& L. Tacconi (Springer: Berlin), p. 173

\reference{}Steidel, C.C.,  Adelberger, Kurt L.,
Shapley, Alice E., Pettini, Max,
Dickinson, Mark, \& Giavalisco, Mauro 2000, ApJ, 532, 170

\reference{}Storrie-Lombardi, L.J., 
McMahon, R.G., Irwin, M.J., \& Hazard, C. 1996, ApJ, 468, 12

\reference{}Strong, A.W., Bloemen, J. B. G. M.,
 Dame, T. M. et al. 1988, A\& A, 201, 1

\reference{}Weiss, A., Neininget, N., H\"uttermeister, S., 
\& Klein, U. 2001, A\& A, 365, 571

\reference{}Wilson, T.L., Serabyn, E., Henkel, C., \&
Walmsley, C.M. 1986, A\& A 158, L1

\reference{}Yun, M.S., Carilli, C.L., Kawabe, R., Tutui, Y., Kohno,
K. \&\ Ohta, K. 2000, ApJ, 528, 171

\end{references}
\end{document}